\documentclass[notitlepage,twocolumn,aps,prl,lengthcheck]{revtex4-2}
\usepackage[utf8]{inputenc}
\usepackage[T1]{fontenc}
\usepackage[english]{babel}
\usepackage{amsmath,amssymb,amsthm}
\usepackage{graphicx}
\usepackage[shortlabels]{enumitem}
\usepackage{xcolor}
\usepackage{csquotes}\MakeOuterQuote"
\usepackage{soul}\setstcolor{red}
\usepackage[colorlinks,citecolor=blue,linkcolor=blue]{hyperref}
\usepackage[capitalize]{cleveref}
\usepackage{comment}

\theoremstyle{definition}

\newtheorem*{problem*}{Problem}
\newtheorem*{assumption*}{Assumption}

\newtheorem*{warning*}{Warning}

\newcommand{\ket}[1]{|#1\rangle}

\newcommand{\bra}[1]{\langle#1|}

\newcommand{\norm}[1]{\lVert #1\rVert}
\newcommand{\oo}{\infty}
\newcommand{\ox}{\otimes}
\newcommand{\mc}{\mathcal}
\newcommand{\eps}{\varepsilon}
\newcommand{\III}{\mathrm{III}}
\newcommand{\II}{\mathrm{II}}
\newcommand{\I}{\mathrm{I}}

\newcommand{\up}[1]{^{(#1)}}

\DeclareMathOperator{\tr}{Tr}

\def\A{{\mc A}}
\def\B{{\mc B}}
\def\CC{{\mathbb C}}

\def\H{{\mc H}}
\def\K{{\mathcal K}}
\def\M{{\mc M}}

\def\O{{\mc O}}

\def\U{{\mc U}}

\def\NN{{\mathbb N}}

\def\ZZ{{\mathbb Z}}

\newcommand*{\1}{\text{\usefont{U}{bbold}{m}{n}1}}
\newcommand{\placeholder}[0]{{\,\cdot\,}}

\begin{document}

\title{Relativistic Quantum Fields Are Universal Entanglement Embezzlers}

\author{Lauritz van Luijk, Alexander Stottmeister,\\ Reinhard F.~Werner, Henrik Wilming}
\affiliation{Institut f\"ur Theoretische Physik, Leibniz Universit\"at Hannover, \\ Appelstraße 2, 30167 Hannover, Germany}
\date{\today}

\begin{abstract}
    Embezzlement of entanglement refers to the counterintuitive possibility of extracting entangled quantum states from a reference state of an auxiliary system (the "embezzler") via local quantum operations while hardly perturbing the latter. We uncover a deep connection between the operational task of embezzling entanglement and the mathematical classification of von Neumann algebras. Our result implies that relativistic quantum fields are universal embezzlers: Any entangled state of any dimension can be embezzled from them with arbitrary precision. This provides an operational characterization of the infinite amount of entanglement present in the vacuum state of relativistic quantum field theories.
\end{abstract}
\maketitle

Entanglement allows a pair of quantum systems to exhibit stronger correlations than classical systems.
Once considered paradoxical, today we think of entanglement as a precious resource that can power quantum information processing. It is an essential ingredient in quantum teleportation \cite{bennett_teleporting_1993} or quantum cryptography \cite{bennett_quantum_1984,ekert_quantum_1991}, and quantum computers require large amounts of entanglement to exhibit an advantage over their classical counterparts \cite{vidal_efficient_2003}. 
In quantum mechanics, various measures of entanglement are available. For example, the entanglement content of a quantum system described by pure states can be quantified using entanglement entropy, which has an operational meaning in terms of entanglement distillation \cite{bennett_concentrating_1996}.

In recent years, increased attention has been paid to entanglement in the context of quantum field theories \cite{srednicki_entropy_1993,holzhey_geometric_1994,calabrese_entanglement_2004,calabrese_entanglement_2009,casini_geometric_2004,ryu_holographic_2006,casini_c-theorem_2007,casini_entanglement_2009,headrick_entanglement_2010,longo_relative_2018,hollands2018entanglement_measures,longo_von_2021,longo_bekenstein-type_2024}. 
It has long been known that quantum states of relativistic quantum fields are highly entangled and that they can be used to violate Bell inequalities perfectly \cite{summers1985vacuum,summers1987,summers_maximal_1988}.
In fact, the entanglement between any region of spacetime and its causal complement is infinite -- a property that asks for an operational explanation.
Building on the holographic principle \cite{hooft_dimensional_1993,susskind_world_1995,maldacena_large-n_1999,witten_anti_1998,bousso_holographic_2002}, it has even been hypothesized that entanglement provides the basis for the geometry of spacetime itself \cite{van_raamsdonk_building_2010,swingle_entanglement_2012,maldacena_cool_2013}.

Like other resources, entanglement deteriorates under usage: Entanglement quantifiers decrease, and the ability of the associated quantum systems to power quantum information processing tasks lessens. 
Like other resources, one is tempted to ask whether entanglement can be extracted in a way that makes it impossible (or arbitrarily hard) to detect -- in other words, to embezzle entanglement.

Here, we study the task of embezzlement of entanglement \cite{van_dam2003universal} and explore its ultimate limits. 
We report on the discovery of a deep link between embezzlement and the mathematical classification of type III von Neumann algebras (see \cite{long_paper} for the proofs of the statements in this paper). 
Our results lend direct operational meaning to the infinite entanglement of quantum states of relativistic quantum fields in terms of embezzlement: 
It is possible to embezzle any quantum state of any dimension to any precision from any quantum state of a relativistic quantum field while making sure that the state of the quantum field is perturbed arbitrarily little. Conversely, the only types of quantum systems that allow for such universal behavior are necessarily of the same mathematical nature as relativistic quantum fields. 
In the following, we explain these results, which can seen as part of a wider effort of exploring the operational significance of the infinite entanglement in quantum systems with infinitely many degrees of freedom by operator algebraic methods (see e.g.  \cite{summers1985vacuum,summers1987,summers_maximal_1988,keyl_infinitely_2003,verch_distillability_2005,Junge2011,cleve_perfect_2017,hollands2018entanglement_measures,crann_state_2020,van_luijk_schmidt_2023,hollands2023channel,long_paper,vanluijk2024critical,vanluijk2024multipartite,vanluijk2024pure}).

\paragraph{Embezzlement of entanglement.} 
Van Dam and Hayden showed for the first time that it is possible to embezzle entanglement if a suitable quantum system is available \cite{van_dam2003universal}.
They proved that the family of pure quantum states 
$\ket{\Omega_n}=C_n  \sum_{j=1}^n \frac{1}{\sqrt{j}} \ket{j}_A\ox\ket{j}_B$
on Hilbert spaces $\mc H\!=\!\mc H_A\!\otimes\!\mc H_B$ of dimension $n^2$ has the following property: For all bipartite quantum states $\ket\Phi,\ket\Psi$ on a $d^2$-dimensional Hilbert space $\mc K\!=\!\mc K_{A'}\!\ox\!\mc K_{B'}$, there exist unitaries $u_{AA'}$ and $u_{BB'}$ acting on $AA'$ and $BB'$, respectively, such that
\begin{align}
        u_{AA'} u_{BB'} \ket{\Omega_{n}}\ox\ket\Phi \approx \ket{\Omega_{n}}\ox \ket\Psi
\end{align}
with an error of $4\frac{\log d}{\log n}$, which can be made arbitrarily small by increasing $n$.
In particular, Alice and Bob can extract arbitrary entangled states $\ket\Psi$ from any unentangled state $\ket\Phi=\ket{\Phi_{A'}}\ox\ket{\Phi_{B'}}$ with arbitrarily small error. It is even guaranteed that any measurement scheme aiming to detect the extraction of entanglement from the system $AB$ has an arbitrarily small success probability.
In short, Alice and Bob have ``embezzled'' entanglement.

On the other hand, for every $n$, there are quantum states $\ket\Psi$ with $n\ll d$ such that the error approaches the maximal error $2$. 

It has been shown that any family of bipartite pure states with the above property must have Schmidt coefficients asymptotically scaling like $1/j$ \cite{leung_characteristics_2014,zanoni_complete_2023}.
Even in an infinite dimensional Hilbert space, one therefore cannot simply take the limit $n\rightarrow \infty$ and obtain a valid quantum state. Indeed, it has been shown before that perfect embezzlement, i.e., without error margin, is incompatible with a tensor product structure \cite{cleve_perfect_2017}.

Besides the foundational importance for our understanding of entanglement, embezzlement also served as an important proof device in quantum information theory, for example, for the celebrated Quantum Reverse Shannon Theorem \cite{Bennett_2014,berta_quantum_2011,Leung2019,coladangelo_two-player_2020} and in the context of non-local games \cite{Leung2013coherent,regev_quantum_2013,cleve_perfect_2017,coladangelo_two-player_2020}.

\paragraph{Commuting operator framework.}
We overcome the hurdles above by relaxing the assumption that $\mc H$ factorizes into $\mc H_A\ox \mc H_B$. Instead, we assume that Alice's and Bob's systems are described by commuting observable algebras $\M_A,\M_B$ acting on a Hilbert space $\mc H$. $\M_{A/B}$ contain the operators that Alice/Bob can use to control and measure their subsystems.

We make the standard assumption that $\mc M_A$ (and similarly $\mc M_B$) is a von Neumann algebra: it is closed in the topology induced by demanding that expectation values $A\mapsto \tr \rho A$ are continuous functions for all $A\in\mc M_A$ and all density matrices  $\rho$ on $\H$.
We still demand that Bob has access to all the operators commuting with Alice's algebra $\mc M_A$, i.e., Bob's algebra $\mc M_B$ is the commutant $\mc M_A'$ of Alice's algebra. 
For simplicity, we assume that Alice's and Bob's parts are purely quantum (no non-trivial classical degrees of freedom), meaning that $\mc M_A$ and $\M_{B}$ are factors: Within $\mc M_{A/B}$, the only operators commuting with all others are scalars.
All these assumptions are met in the usual tensor product scenario by setting $\M_A = \B(\H_A)\ox\1_B$, $\M_B=\1_A\ox \B(\H_B)$. 
Embezzlement of entanglement is expressed in this setting by the existence of a vector $\ket{\Omega}\in\H$ such that for arbitrary target states $\ket{\Phi},\ket{\Psi}\in\mc K_{A'}\ox\mc K_{B'}$ and error threshold $\eps>0$:
\begin{align}\label{eq:mbz_bipartite}
    \norm{ u_{AA'} u_{BB'}\ket{\Omega}\ox \ket{\Phi} - \ket{\Omega}\ox\ket\Psi} < \eps,
\end{align}
for local unitaries $u_{AA'} \in\M_{A}\otimes\mc B(\mc K_{A'})\ox\1_{B'}$, $u_{BB'}\in\M_{B}\ox \1_A\otimes\mc B(\mc K_{B'})$ of Alice and Bob, respectively.
\begingroup
\setlength{\tabcolsep}{10pt} 
\renewcommand{\arraystretch}{1.5} 
\begin{table}[t!]
    \centering
    \begin{tabular}{c|ccccc}
    type & I& II& III$_0$& III$_\lambda$&  III$_1$\\\hline\hline
    $\kappa_{\textit{min}}$&$2$&$2$&$0$ or $2$&$0$&$0$\\
    $\kappa_{\textit{max}}$&$2$&$2$&$2$&$2 \frac{1-\sqrt{\lambda}}{1+\sqrt{\lambda}}$&$0$
    \end{tabular}
    \caption{Values of the invariants $\kappa_{\textit{min}}$ and $\kappa_{\textit{max}}$ for factors of a given type (with $0<\lambda <1$). 
    Since $2\frac{1-\sqrt\lambda}{1+\sqrt\lambda}$ is bijective in $\lambda$, the subtype of a type III factor can be determined from its embezzling capability. See \cite{long_paper} for full details and rigorous proofs.}
    \label{table:kappa}
\end{table}
\endgroup
Since the situation is completely symmetric between Alice and Bob, we subsequently drop the subscripts and simply write $\mc M$ and $\mc M'$. Moreover, we can invoke this symmetry to phrase the bipartite statement \cref{eq:mbz_bipartite} solely in terms of Alice's system.

\paragraph{Monopartite vs bipartite embezzlement.}
Just as transformations between entangled pure states via local operations and classical communication (LOCC) can be reduced to studying their marginals on one system \cite{nielsen_conditions_1999}, embezzlement of entanglement can be reduced to \emph{monopartite embezzlement} of marginals (reduced states) on Alice's systems. A quantum state $\omega$ on a von Neumann algebra $\M$ assigns expectation values to operators via $A\mapsto \omega(A)$ and may be represented by a density matrix on $\H$ so that $\omega(A)=\tr\rho A$. 
Since different density matrices may induce the same state on $\M$, quantum states are described by (positive and unital) linear functionals on $\M$ and not by density operators on $\H$.
Suppose now that for some state $\omega$ on $\mc M$ and a unitary $u\in\mc M\ox\mc B(\mc K)$ we have
\begin{align}\label{eq:mbz_single}
    \norm{ u (\omega \ox \varphi )u^* -  \omega\ox \psi} <\varepsilon^{2},
\end{align}
where $\varphi$ and $\psi$ are (generally mixed) states on $\mc B(\mc K)$, and $\|\cdot\|$ is the norm on the dual space of $\M\ox\B(\K)$.
Then, our assumptions imply (see \cite{long_paper}) that there exists a unitary $u'\in\mc M'\ox\mc B(\mc K')$ such that \cref{eq:mbz_bipartite} holds for purifications $\ket\Phi,\ket \Psi\in \mc K\ox \mc K'$ of $\varphi$ and $\psi$, and $\ket\Omega\in\H$ of $\omega$, respectively. Therefore monopartite embezzlement in the sense of \eqref{eq:mbz_single} implies embezzlement of entanglement (in the sense of \eqref{eq:mbz_bipartite}).
Conversely, it is clear that embezzlement of entanglement implies monopartite embezzlement by reducing to Alice's systems. Thus, the two are equivalent.

We can measure how well a state $\omega$ on $\mc M$ can embezzle arbitrary finite-dimensional quantum states by the quantity
\begin{align}\label{eq:mbz_quant}
\kappa(\omega) := \sup_{d} \sup_{\psi,\,\varphi} \inf_{u} \,\norm{u (\omega\ox \varphi) u^* - \omega\ox\psi}.
\end{align}
Here, $\psi,\varphi$ are states on an $d$-dimensional Hilbert space $\mc K$ and $u \in \mc M\ox \mc B(\mc K)$ is unitary. 
The derived quantities
\begin{align}
    \kappa_{\textit{min}}(\mc M) := \inf_\omega \kappa(\omega),\quad \kappa_{\textit{max}}(\mc M) := \sup_\omega \kappa(\omega), 
\end{align}
where we optimize over states on $\M$, quantify the best and worst possible embezzling performances of states on $\mc M$.

$\kappa_{\textit{min}}(\mc M)$ and $\kappa_{\textit{max}}(\mc M)$ are algebraic invariants of $\mc M$, which allow us to classify von Neumann algebras. To state our main technical result, we recall that there is a standard classification of factors into types I, II, and III. Famously, Alain Connes provided a finer classification of type III algebras into subtypes III$_{\lambda}$ with $\lambda\in[0,1]$ using deep arguments based on modular theory \cite{connes1973classIII,connes1976injective}.
Below and in the end matter we provide examples of factors of the different types in terms of infinite spin chains.
With this in mind, our discovery is that the derived invariant $\kappa_{\textit{max}}$, and hence embezzlement, precisely recovers the subtypes of Connes' classification of type III factors as shown in Tab.~\ref{table:kappa}.
These results show that all type III$_\lambda$ factors with $\lambda>0$ admit an \emph{embezzling state}: a state that can embezzle any pure, bipartite quantum state of arbitrary dimension with arbitrary accuracy while being disturbed arbitrarily little.
In particular, we find that the worst error $\kappa_{\textit{max}}(\mc M)$ strictly decreases as $\lambda\rightarrow 1$.
Most importantly, type III$_1$ factors are characterized by the fact that \emph{every} state is an embezzling state. Hence, we call systems described by type III$_1$ factors \emph{universal embezzlers}, exhibiting the strongest form of infinite entanglement quantified by $\kappa_{max}$.
In contrast, type I and II algebras admit no embezzling states -- there always exist pairs of states $\varphi,\psi$ that lead (arbitrarily close) to the maximally allowed value $\kappa(\omega)=2$.
Remarkably, there is a strict dichotomy: A quantum system either has an embezzling state, or it does not admit any form of approximate embezzlement in the sense that $\kappa(\omega)=2$ for all states.
Therefore, the invariant $\kappa_{\textit{min}}$ can only take on the extremal values $0$ or $2$.

\paragraph{Techniques.}
We briefly sketch how the results are obtained:
\begin{figure}
 	\centering
 	\includegraphics[width=8.5cm]{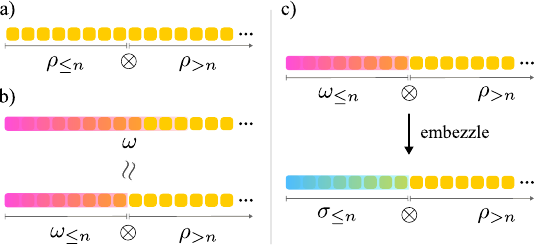}
 	\caption{{\bf Why universal embezzlers have type III$_1$.} a) An infinite spin chain is described relative to a product state $\rho = \otimes_{j=1}^\infty \rho_j$. b) The states of the spin chain can be arbitrary on the first $n$ spins, but converge to $\rho$ on the remaining spins. This allows to approximate any state $\omega$ by $\omega_{\leq n}\ox \rho_{>n}$ for sufficiently large $n$. c) If the spin chain is a universal embezzler $\rho_{>n}$ must already be an embezzling state. We can hence (approximately) unitarily transform $\omega\approx \omega_{\leq n}\ox \rho_{>n}$ to $ \sigma_{\leq n}\ox \rho_{>n}\approx \sigma$, for any two states $\omega,\sigma$.}
 	\label{fig:ITPFI}
\end{figure}
By utilizing the "flow of weights" introduced by Connes and Takesaki \cite{connes1973classIII, takesaki_duality_1973, connes1977flow}, one can associate to a state $\omega$ on a von Neumann algebra $\M$ a probability distribution $P_\omega$ on a classical dynamical system \cite{haagerup_standard_1975}. The latter will be ergodic if $\M$ is a factor.
Building upon work by Haagerup and St\o rmer in \cite{haagerup1990equivalence}, we show that the embezzlement quantifier $\kappa(\omega)$ measures precisely how much $P_\omega$ deviates from being invariant under the ergodic flow -- with embezzling states corresponding to invariant distributions. We can calculate $\kappa$, $\kappa_{\textit{min}}$ and $\kappa_{\textit{max}}$ by studying the classical ergodic system associated with a factor.

\paragraph{Universal embezzlers.}
In the following, we explain why universal embezzlers are precisely described by type III$_{1}$ factors without using the machinery from the previous paragraph:
Such factors are uniquely characterized by the \emph{homogeneity of the state space}, discovered by Connes and St\o rmer \cite{connes_homogeneity_1978}: For any two states $\omega_1,\omega_2$ and every $\eps>0$ there exists a unitary in $\M$ such that
\begin{align}
    \norm{u\omega_1 u^* - \omega_2} < \eps.
\end{align}
Another crucial property is that $\M \cong \M\ox \mc B(\mc K)$ for any $n$-dimensional Hilbert space $\mc K$.
Combining the latter property with the homogeneity of the state space tells us that we can find a unitary $u$ for any state $\varphi$ on $\mc B(\mc K)$ and any precision $\varepsilon>0$ such that $u\omega u^* \approx_\eps \omega\ox \varphi$.   

As a converse, we provide a heuristic argument for factors appearing in many-body physics and quantum field theory to be of type III$_1$ if they are universal embezzlers (see Fig.~\ref{fig:ITPFI} for an illustration and \cite{long_paper} for a proof):
The set-up is as follows: The von Neumann algebras arising in physics allow for finite-dimensional approximations (i.e., are `hyperfinite'). We can view any such algebra as the von Neumann algebra $\mc M$ of a half-infinite chain of uncorrelated spin-1/2 particles in a product state $\rho = \otimes_{j}\rho_{j}$ \footnote{The only possible exceptions are type $\III_{0}$ algebras, which are uncommon in physics applications.}.
Such spin chains enjoy two important properties:

First, we can remove any finite number of spins from the spin chain, with the remainder being unitarily equivalent to the original. 
Second, all states $\omega$ on $\M$ can be approximated to arbitrary precision $\varepsilon>0$ as
\begin{align}
    \omega \approx_{\varepsilon} \omega_{\leq n}\ox \rho_{> n},\quad \rho_{> n} := \otimes_{j=n+1}^\infty \rho_j,
\end{align}
for sufficiently large $n$, where $\omega_{\leq n}$ is the restriction of $\omega$ to the first $n$ spins. Informally, we may say that all states "agree at infinity" with the product state $\rho$.

Given the above, the heuristic argument goes as follows: Suppose that $\mc M$ is universal embezzling.
The first property tells us that if every state on $\mc M$ is embezzling, the same will hold for every state after removing the first $n$ spins.
In particular, the states $\rho_{> n}$ are embezzling states. Since the first $n$ spins are described by a finite-dimensional Hilbert space, we find that for any state $\omega'_{\leq n}$ on the first $n$ spins and any precision $\eps>0$, there is a unitary $u$ such that
\begin{align}
    u(\omega_{\leq n}\ox \rho_{>n})u^* \approx_{\varepsilon} \omega'_{\leq n}\ox\rho_{>n}. 
\end{align}
By the second property, all states are of this form to arbitrary precision. Hence any two states $\omega,\omega'$ on $\mc M$ are approximately unitarily equivalent: $u\omega u^* \approx \omega'$, i.e., $\M$ has a homogeneous state space and must be of type III$_1$.

\paragraph{Embezzlement and relativistic quantum fields.}
In the algebraic approach to relativistic quantum field theory, one associates to every open region $\O$ in spacetime a von Neumann algebra $\M(\O)$ capturing the properties of quantum fields (localized in $\O$), e.g., operators representing the local field strength of the electromagnetic field.
In the following, we give a brief overview of the operator-algebraic structure of QFT.
All algebras act jointly irreducibly on a common separable Hilbert space $\mc H$ with the (unique) vacuum state described by a vector $\Omega\in\mc H$. Local consistency is encoded by demanding that the local algebra of a region $\O_{A}\cup\O_{B}$ is generated by $\mc M_A = \M(\O_{A})$ and $\mc M_B = \M(\O_{B})$, denoted by $\M_{A}\vee\M_{B}$.
Einstein locality (or causality) is implemented by the relative commutativity of local algebras $\mc M_A$ and $\mc M_B$ for spacelike separated regions $\O_A$ and $\O_B$. Completeness of relativistic dynamics \footnote{This is also sometimes called the Time-Slice Axiom.} is realized by $\mc M(\O'') =\mc M(\O)$, where $\O''$ the causal completion (or Cauchy development) of $\O$. 

\begin{figure}[t]
 	\centering
 	\includegraphics[width=\linewidth]{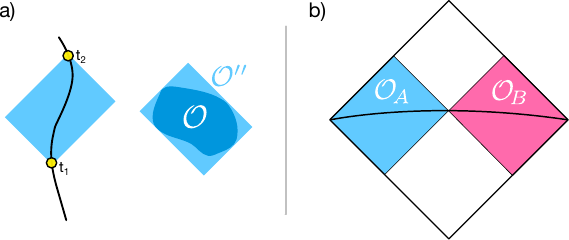}
 	\caption{{\bf Embezzling from Quantum Fields.} a) Left: The spacetime region accessible to an observer between two events $A,B$ on his worldline. Right: The causal closure $\mathcal O''$ of some open subset $\mathcal O$ of spacetime. b) Penrose diagram of Minkowski space. Alice and Bob have access to the left wedge $\mathcal O_A$ and the right wedge $\mathcal O_B = \mathcal O_A'$, respectively. The curved horizontal line corresponds to an equal-time slice.}
 	\label{fig:QFT}
\end{figure}

Consider now an open set $\O_A$ and its causal complement $\O_B=\O_A'$. 
For instance, $\O_A$ and $\O_B$ could be complementary wedges $\O_{A/B} = \{ x^\mu : |x^0|< \mp x^1\}$ in Minkowski space (see Fig.~\ref{fig:QFT}).
Then, it is meaningful to consider $\H$ with $\mc M_A$ and $\mc M_B$ as a bipartite system, where Alice has access to the spacetime region $\O_A$, and Bob has access to the spacetime region $\O_B$.
In this setting, the Bisognano-Wichmann theorem \cite{bisognano1976duality} and the Reeh-Schlieder theorem \cite{reeh_bemerkungen_1961,haag_local_1996} guarantee that $\M_{A}'=\M_{B}$ (Haag duality).

Since our basic assumptions for embezzlement are fulfilled, we can ask how well the vacuum (or any other state) on $\mc M=\mc M_A$ performs at embezzling.
By the above, this task amounts to determining the type of the von Neumann algebra $\mc M$. 
This problem has been studied intensely before; see \cite{haag_local_1996,yngvason_role_2005} for an overview of results. 
It has been found under very general conditions, irrespective of whether the quantum fields are interacting or not, that the algebras $\mc M(\O)$ have type III$_1$, including the situation of so-called wedge algebras as considered above \cite{driessler1975lightlike, driessler1977type, longo1979notes, fredenhagen_modular_1985, buchholz_universal_1987}.
We conclude that relativistic quantum fields are universal embezzlers:

\emph{From any quantum state and any bipartition of Minkowski space as above, Alice and Bob can embezzle any finite-dimensional quantum state to any precision they desire.}

\paragraph{Discussion and outlook.} It is often emphasized that the vacuum of relativistic quantum field theories is "infinitely entangled". 
This infinite entanglement is difficult to interpret physically.
As the classification of von Neumann algebras shows, there can be different types of "infinite entanglement," but only type III$_1$ allows for embezzlement from any state (see Tab.~\ref{table:kappa}).
Our result provides a direct operational and quantitative charaterization of the fact that the local algebras of relativistic quantum field theories are of type III$_{1}$.
As mentioned in the introduction, it has long been known that the vacuum of a relativistic quantum field can be used to violate Bell inequalities arbitrarily well. Our result makes transparent why this is the case: 
Alice and Bob could simply embezzle a spin-1/2 Bell state and then apply a Bell test \cite{bell_einstein_1964,clauser_proposed_1969}. 

A natural question is whether the embezzlement quantifier $\kappa(\omega)$ is monotone under local operations and classical communication (LOCC).
This question is nontrivial only if $\kappa$ can take on different values, requiring $\M$ to have type $\III$ (see Tab.~\ref{table:kappa}).
Surprisingly, under an LOCC paradigm for general von Neumann algebras \cite{vanluijk2024pure}, all pure states are equivalent in the type $\III$ case, implying that $\kappa$ cannot be monotone.
We caution that these results need to be taken with a grain of salt when applied to quantum field theory as it is a currently much-debated issue how to properly make sense of general local operations \cite{fewster2020local_measurements}.

A concern regarding quantum fields is the localizability of the unitaries applied by Alice and Bob. 
In our discussion, Alice and Bob seem to require access to their whole patches of spacetime to embezzle arbitrary quantum states. However, for any finite error $\eps$ in \eqref{eq:mbz_bipartite}, Alice and Bob only need to act on finite regions, which can moreover be properly space-like separated. As $\eps\rightarrow 0$, the regions grow, and their separation decreases. See the End Matter and \cite{vanluijk2024multipartite} for a detailed explanation. 
The split property of quantum field theory shows that the infinite amount of entanglement in quantum fields is localized at the boundary between two regions of spacetime \cite{buchholz1974product_states, doplicher1984split, buchholz1986causal_independence}. 
This suggests that Alice and Bob may, in fact, only have to act close to their common boundary to embezzle quantum states \cite{summers_maximal_1988}. 
Proving this will require more specific properties of quantum fields than the general treatment given here. 
It would also be interesting to see explicitly worked-out unitaries in a free field. 
We leave these problems open for the future. 

It has recently been argued that, in the presence of gravitation, local algebras of observables are of type II instead of type III$_1$. Specifically, type II$_{\infty}$ was found outside of a blackhole horizon \cite{witten2022crossed} for the observables of a single observer, while II$_{1}$ was found for static de-Sitter patches \cite{chandrasekaran2023dsobs} (see also \cite{longo2022note, chandrasekaran2023adscft}) -- both for single observers and for certain bipartite situations involving spacelike separated observers. 
As a consequence, gravitation might not only disable universal embezzlement but even prohibit the existence of embezzlers altogether, rendering the possibility of embezzlement a distinguishing feature between ordinary relativistic quantum field theory and potential theories of quantum gravity. In \cite{vanluijk2024pure}, we have studied pure state entanglement for general factors to clarify the differences pertaining to the various types.

Our work also provides an operational motivation for studying the different classes of infinite entanglement arising in ground states of quantum many-body systems (see \cite{matsui_split_2001,matsui_boundedness_2013,keyl2006,naaijkens_localized_2011,naaijkens_haag_2012,naaijkens_kitaevs_2015,fiedler_haag_2015,naaijkens_quantum_2017,moon_automorphic_2019,ogata_type_2022,ogata_derivation_2022,jones_local_2023}
for some prior results). In \cite{vanluijk2024critical}, we have shown that ground state sectors of one-dimensional, translation-invariant, critical, quasi-free fermion systems and their associated spin chains are universal embezzlers, showing that embezzlement naturally appears in many-body physics. Additionally, we identified a criterion providing a one-to-one correspondence between embezzling families and embezzling states that extends to the multipartite setting in \cite{vanluijk2024multipartite}. This criterion connects the embezzling capabilities of many-body ground states in finite volume to those of various limiting situations, e.g., infinite-volume or scaling limits \cite{stottmeister2020oar,vanluijk2024convergence}. 
Finally, in \cite{vanluijk2024multipartite}, we also constructed the first \emph{multipartite} embezzling state. The construction raises the interesting question of whether vacua of relativistic quantum fields or ground states of quantum many-body systems can also be multipartite embezzlers.\\

\begin{acknowledgments}{\it Acknowledgments.}
    This work was motivated by unfinished prior work of RFW together with Volkher Scholz and Uffe Haagerup. We thank Marius Junge, Roberto Longo, Yoh Tanimoto, and Rainer Verch for useful discussions. LvL and AS have been funded by the MWK Lower Saxony via the Stay Inspired Program (Grant ID: 15-76251-2-Stay-9/22-16583/2022).
\end{acknowledgments}

\onecolumngrid
\bigskip
\section{End Matter}
\twocolumngrid

\paragraph{Quantum information and von Neumann algebras.}
As explained in the main text, the mathematical description of physical systems with infinitely many degrees often requires an algebraic approach. The role of the observable algebra, in traditional quantum mechanics the algebra $\B(\H)$ of all bounded operators on a Hilbert space $\H$, is taken over by a more general von Neumann algebra $\M$ (for a general introduction, see \cite{takesaki1}, and \cite{bratteli1987oa1} from a mathematical physics point of view). 
The observables, in Quantum Information only required to be positive operator-valued measures, then are positive $\M$-valued measures. 
There are two ways to define von Neumann algebras \cite{sakai}: The first takes them concretely as subalgebras of $\B(\H)$ for some Hilbert space $\H$, which are closed in the weak operator topology (defined as the topology making matrix elements $A\mapsto\bra{\phi}A\ket\psi$ continuous).  This view arises naturally  when the system under consideration is a {\it sub}system of a standard quantum system with observable $\B(\H)$. The second definition is more abstract, namely as an algebra with adjoint operation and norm satisfying $\norm{a^*a}=\norm a^2$ (i.e., a so-called C*-algebra) which, as a normed space, is the dual of some Banach space $\M_*$. These definitions are equivalent in that every algebra satisfying the abstract definition is isomorphic to one of the concrete sort. The link between these is the notion of {\it normal states}: In the concrete case, these are the states (probability functionals) arising as the trace with a density operator. In the abstract case, these are the probability functionals defined by the pre-dual $\M_*$, also characterized as the states which are continuous under limits of increasing bounded nets in $\M$. Intuitively, normal states are those that can be prepared  relatively easily, for example, requiring only finite energy.  For example, in ordinary quantum mechanics, states with sharp, pointlike position distribution require infinite momentum and, hence, infinite kinetic energy. Accordingly, all normal states (density operators) produce position distributions, which have a density with respect to the Lebesgue measure.

Given two von Neumann algebras $\mc M_A$ and  $\mc M_B$ contained in an ambient system with observable algebra $\B(\H)$, the condition that all observables of $\mc M_A$ can be measured jointly with all those of $\mc M_B$, is equivalent to the vanishing of all commutators  $[a,b]=0$ for $a\in\mc M_A$ and  $b\in\mc M_B$, i.e., we are in the commuting operator framework described in the text. A compact notation for this is $\mc M_B\subset\mc M_A'$, which uses the {\it commutant} $\mc R'=\{x\in\B(\H)|\forall_{r\in\mc R} [x,r]=0\}$ for a subset $\mc R\subset\B(\H)$. The {\it center} of a von Neumann algebra $\mc M$ is $\M\cap\M'$ and is the classical subsystem consisting of those observables, which can be measured together with all others or, equivalently, essentially without disturbance. If the classical part is trivial, i.e., $\M\cap\M'=\CC\1$, one calls $\M$ a {\it factor}, or purely quantum.  In adversarial contexts, e.g., in cryptography, it is assumed that the adversary, e.g., the eavesdropper, can measure everything that can possibly be measured without disturbing the system. That is, we assume that her observable algebra is $\M'$. This results in bipartite systems with $\M_A=\M$ and $\M_B=\M'$, which are assumed above. That this is a symmetric relation between $A$ and $B$ follows from von Neumann's ``Bicommutant Theorem'': Each von Neumann algebra $\M\subset\B(\H)$ is equal to its second commutant $\M'':=(\M')'=\M$. In this sense, the commutant is a complement for factors: $\M$ and $\M'$ have trivial intersection, but together generate the ambient system $\B(\H)$. 

When $\M\subset\B(\H)$ is a factor,  operations  that are local to this subsystem can be characterized by the condition that they have an extension to $\B(\H)$ which maps normal states to normal states and leaves the commutant $\M'$, i.e., the observables of a potential adversary, invariant. This is equivalent to a Kraus form $a\mapsto\sum_ik_i^*ak_i$ with $k_i\in\M$. Again, equivalently, we can implement the operation on the whole system by coupling it to an ancilla with a tensor product, engineering an interaction using only operators from $\M$ and the ancilla, and tracing out the ancilla. Reversible local operations are thus implemented as $a\mapsto u^*au$ with unitaries $u\in\M$. This form is used in the main text.

In general, a physical system can have both classical and quantum degrees of freedom. 
This is reflected by the fact that the von Neumann algebra $\M$ decomposes (uniquely) as a direct integral $\M = \int^\oplus_\Gamma \M_\gamma \,d\mu(\gamma)$ of factors $\M_\gamma$ (describing purely quantum systems), over its center (describing all classical degrees of freedom) $\M\cap \M' = L^\oo(\Gamma,d\mu)$ \cite[Sec.~IV.8]{takesaki1}.
Thus, understanding the different observable algebras appearing in physical systems amounts to a mathematical classification of all possible factors.
Essentially all factors appearing in physics are `approximately finite-dimensional', and these have been classified by the collected works of many (in particular, Connes \cite{connes1973classIII} and Haagerup \cite{haagerup_uniqueness_1987}).
Factors come in three types, surprisingly called I, II, and III.
Type I factors are finite or infinite matrix algebras describing "ordinary quantum mechanics". Types II  and III are more wild objects describing systems in many-body physics and field theories. Type III factors, playing an important role in the main text, are classified into subtypes III$_\lambda$, $0\le\lambda\le 1$.
In the following, we give concrete examples of factors arising in physics:

We consider the spin-chain C*-algebra $\bigotimes_{\ZZ}M_{2}(\CC)$, which will naturally give rise to a bipartite system if split into half-chain algebras $\A=\bigotimes_{\ZZ\setminus\NN}M_{2}(\CC)$ and $\B=\bigotimes_{\NN}M_{2}(\CC)$. 
The various types of factors arise from simple product state constructions for spin chains (so-called Powers and Araki-Woods factors \cite{powers1967uhf,araki1968factors}).
On the full spin chain, we consider the (bipartite) pure states
\begin{align}\label{eq:powers}
\ket{\Omega_{\lambda}} & \!\!=\!\!\bigotimes_{j=1}^{\infty}\!\tfrac{1}{\sqrt{1+\lambda_{j}}}\!\Big(\!\ket{1}_{\!-j+1}\!\!\otimes\!\ket{1}_{\!j}\!+\!\!\sqrt{\!\lambda_{j}}\ket{2}_{\!-j+1}\!\!\otimes\!\ket{2}_{\!j}\!\Big),
\end{align}
where $\lambda_{j}\in[0,1]$ and $\ket{k}_{j}$ refers to the $k$th standard basis vector at site $j$. It is known that the reduced states $\omega_{\lambda}|_{\A/\B} = \bra{\Omega_{\lambda}}(\placeholder)_{\A/\B}\ket{\Omega_{\lambda}}$ give rise to factors, $\M_{A/B} = \pi_{\omega_{\lambda}}(\A/\B)''$, of all possible types, depending on local Schmidt spectra $\lambda_{j}$. For a complete statement about the various types, we refer to the seminal paper of Araki and Woods \cite{araki1968factors}. For the specific case that the local Schmidt spectra are constant along the chain, i.e., $\lambda_{j}=\lambda$, the resulting types of factors are as follows: $\M_{A/B}$ is of type $\I_{\infty}$ if $\lambda=0$, of type $\II_{1}$ if $\lambda=1$, and of type $\III_{\lambda}$ else.

Another physically motivated example is given by the ground state $\ket{\Omega_{(\gamma,h)}}$ of the $XY$ Hamiltonian
\begin{align}\label{eq:xy}
H & \!=\!-\!\sum_{j\in\ZZ}\!\big(\!\tfrac{1+\gamma}{2}\sigma^{x}_{\!j}\sigma^{x}_{\!j+1}\!+\!\tfrac{1-\gamma}{2}\sigma^{y}_{\!j}\sigma^{y}_{\!j+1}\!+\!h\sigma^{z}_{\!j}\big),
\end{align}
with transverse magnetic field $h$ and anisotropy $\gamma$. In the isotropic case, $\gamma=0$, with sufficiently small magnetic field $|h|<1$, the half-chain von Neumann algebras are of type $\III_{1}$ \cite{matsui_split_2001,keyl2006,vanluijk2024critical}.

\null
\paragraph{Localization of embezzlement in quantum field theory.}

We explain why Alice and Bob can localize their embezzling unitaries $u_{AA'}$ and $u_{BB'}$, respectively, in the quantum field theory setting discussed in the main text (abstract versions of this argument are given in \cite{vanluijk2024critical,vanluijk2024multipartite}).

\begin{figure}[ht!]
    \centering
    \includegraphics[width=6cm]{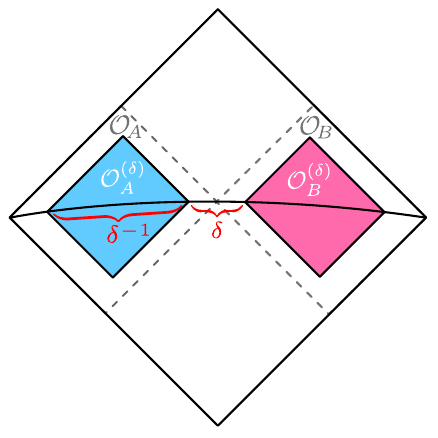}
    \caption{{\bf Localized Embezzlement from Quantum Fields.} A Penrose diagram of Minkowski space. Alice and Bob have access to the left wedge $\mathcal O_A$ and the right wedge $\mathcal O_B = \mathcal O_A'$, respectively. The curved horizontal line corresponds to an equal-time slice. $\O_{A}\up\delta$ and $\O_{B}\up\delta$ are $\delta$-local observable algebras that allow for embezzling given initial and final states up to some fixed error threshold $\eps = \eps(\delta)>0$.}
    \label{fig:local_diamonds}
\end{figure}

We consider the bipartite system where Alice and Bob control the von Neumann algebras $\M_{A}=\M(\O_{A})$ and $\M_{B}=\M(\O_{B})$ associated with complementary wedges $\O_{A/B} = \{ x^\mu : |x^0|< \mp x^1\}$ in Minkowski space (see Fig.~\ref{fig:QFT}).
We may generate the algebras $\M_{A/B}$ of the full wedges by increasing families of the local algebras $\M^{(\delta)}_{A/B} = \M(\O_{A/B}\up\delta)$ of causal diamonds such that $\O_{A}\up\delta$ and $\O_{B}\up\delta$ are spacelike separated on the order of some length scale $\delta$. At the same time, each diamond has a base $\B_{A/B}\up\delta$ with a diameter of the order $\delta^{-1}$ (see Fig.~\ref{fig:local_diamonds}):
\begin{align}\label{eq:eps_diamonds}
    \M_{A/B} & = \Big(\bigcup_{\delta>0}\M_{A/B}^{(\delta)}\Big)''.
\end{align}
Concretely, we may choose causal diamonds over spatial bases $\B_{A/B} = B_{1/\delta}(\pm (1/\delta+\delta/2,0,0))$, where $B_r(\vec x)$ denotes the ball with radius $r$ centered at $\vec x$ in the time-zero plane (see Fig.~\ref{fig:local_diamonds}).
It is a general fact that any unitary in $\U(\M_{A/B}\ox\B(\K_{A'/B'}))$ can be approximated (in the strong* topology) by unitaries in $\bigcup_{\delta\in(0,1]}(\M_{A/B}^{(\delta)}\ox\B(\K_{A'/B'}))$ (see \cite[Lem.~19]{vanluijk2024multipartite}).
Now, given an error $\eps>0$, initial and final states $\ket{\Psi},\ket{\Phi}\in\K_{A'}\otimes\K_{B'}=\K$ and embezzling unitaries $u_{AA'}$ and $u_{BB'}$ such that \cref{eq:mbz_bipartite} holds (with $\tfrac{\eps}{3}$), we can find ($\delta$-local) approximations $u^{(\delta)}_{AA'/BB'}\in\M_{A/B}^{(\delta)}\ox\B(\K_{A'/B'})$ such that $\norm{ (u^{(\delta)}_{AA'/BB'} - u_{AA'/BB'})\ket{\Omega}\ox \ket{\Phi} }<\tfrac{\eps}{3}$ entailing 
\begin{align}\label{eq:local_approx}
    \norm{ u^{(\delta)}_{AA'} u^{(\delta)}_{BB'}\ket{\Omega}\ox \ket{\Phi} - \ket{\Omega}\ox\ket\Psi} & < \eps,
\end{align}
for sufficiently small $0<\delta$. Note that $\delta$ only depends on $\eps$ and the dimension of $\K$.
Thus, Alice and Bob may localize their embezzling unitaries for a fixed error threshold and given initial and final states.

\clearpage

\end{document}